\documentstyle[letter,epsf]{ptptex}
\title {The Cosmic QCD Phase Transition, Quasi-baryonic Dark Matter and
Massive Compact Halo Objects}
\author{Shibaji Banerjee\thanks{Electronic Mail :
phys@bosemain.boseinst.ac.in}\(^{a),b)}\),
Abhijit Bhattacharyya\thanks{Electronic Mail : 
abhijit@th.physik.uni-frankfurt.de}\(^{c),d)}\),
Sanjay K. Ghosh\thanks{Electronic Mail :
sanjay@bosemain.boseinst.ac.in}\(^{a)}\),\\ 
Sibaji Raha\thanks{Electronic Mail :
sibaji@bosemain.boseinst.ac.in}\thanks{Electronic Mail :
raha@rcnp.osaka-u.ac.jp}\(^{a),e)}\),
Bikash Sinha\thanks{Electronic Mail : bikash@veccal.ernet.in}\(^{f),g)}\) and 
Hiroshi Toki\thanks{Electronic Mail : toki@rcnp.osaka-u.ac.jp}\(^{e)}\)}
\inst{ \(^{a)}\) Physics Department, Bose Institute,
93/1, A.P.C.Road, Kolkata 700009, INDIA \\
\(^{b)}\) Physics Department, St. Xavier's College, 30,
Park Street, Kolkata 700016, INDIA \\
\(^{c)}\) Physics Department, Scottish Church College,
3 Cornwallis Square, Kolkata 700006, INDIA \\
\(^{d)}\) Institut f\"ur Theoretische Physik, Universit\"at Frankfurt, 
Robert-Meyer-Str. 8-10, D-60054 Frankfurt-am-Main, GERMANY \\
\(^{e)}\) Research Center for Nuclear Physics, Osaka
University, Mihogaoka 10-1, Ibaraki City, Osaka 567-0047, JAPAN \\
\(^{f)}\) Saha Institute of Nuclear Physics, 1/AF,
Bidhannagar, Kolkata 700064, INDIA \\
\(^{g)}\) Variable Energy Cyclotron Centre, 1/AF, Bidhannagar, Kolkata
700064, INDIA \\}
\abst{We propose that the cold dark matter (CDM) is composed entirely of 
quark matter, arising from a cosmic quark-hadron transition. We denote 
this phase as "quasibaryonic", distinct from the usual baryons. We show 
that compact gravitational lenses, with masses around 0.5 \( M_{\odot} \), 
could have evolved out of the quasibaryonic CDM.}  
\recdate{}
\begin{document}
\maketitle
\par
The present consensus in cosmology is that the universe is flat (\( \Omega 
\sim \) 1), the baryons contributing only about 10\% of the total energy
({\it i.e.} \( \Omega_B \sim \) 0.1). There is an abundance of other matter, 
the {\it cold dark matter} (CDM), which accounts for clumping on small
(galactic/supergalactic scales), amounting to \( \Omega_{CDM} \sim \) 0.35.
The rest of the closure density arises from some kind of vacuum energy, as
yet very poorly understood and termed dark energy, which is believed to be
responsible for the accelerated expansion of the universe. (We shall discuss 
the possible relevance of the scenario described in this letter to dark 
energy on a separate occasion.) The estimate of \( \Omega_B \) comes from 
Big Bang Nucleosynthesis (BBN), whose success is one of the basic tenets of
the standard cosmological model. It is thus believed that the CDM is 
nonbaryonic; speculations about the nature of CDM, all essentially beyond
the standard model of particle interactions (QCD and Electroweak), abound 
and search for these exotic particles is a most active field of research. 
\par
In recent years, there has been experimental evidence\cite{rf:1,rf:2} for 
at least one form of dark matter - the Massive Astrophysical Compact Halo 
Objects (MACHO) - detected through gravitational microlensing 
effects.\cite{rf:3} \ Based on about 13 - 17 Milky Way halo MACHOs detected 
in the direction of LMC - the Large Magellanic Cloud, a Bayesian analysis 
yields their mass estimate in the  (0.15-0.95) \( M_{\odot} \), with the
most probable value being  0.5 \( M_{\odot} \)\cite{rf:4,rf:5}, substantially 
higher than the fusion threshold of 0.08 \( M_{\odot} \). (It is thus 
hard to understand why they do not ignite.) \ The MACHO 
collaboration\cite{rf:4} suggests that the lenses are in the galactic halo. 
In such a case, the observed number of events is already too large to be
reconciled with the BBN limit of \( \Omega_B \), even if the lenses are 
thought to be as faint as blue dwarfs. There are alternate attempts at 
circumventing this difficulty, which include thick disk halo model, tidal wave
debris, self-lensing of the LMC and many others; all of them cannot be 
adequately mentioned in this letter. We adopt the viewpoint, in line with 
the claim of the MACHO collaboation, that the lenses are indeed in the 
galactic halo. In such circumstances, MACHOs cannot be composed of normal
baryons for reasons mentioned above. There have, however, been 
suggestions\cite{rf:6,rf:7} that they could be primordial black holes (PBHs) 
( \( \sim \) 1 \( M_{\odot} \) ), arising from horizon scale fluctuations.
This suggestion requires a fine tuning of the initial density perturbation 
and has been criticised in the literature.\cite{rf:8} 
\par
Within the lore of the standard model, there occurred a phase transition 
from the quark-gluon phase to the hadronic phase during the microsecond 
era after the initial Big Bang, at a temperature of \( \sim \) 100 MeV.
The order of this phase transition is still unsettled;\cite{rf:9} \ lattice 
calculations suggest that in a pure ({\it i.e.} only gluons)  \( SU(3) \) 
gauge theory, it is of first order. In the presence of dynamical quarks on 
the lattice, the situation is more complicated. However, in the early universe,
the large size of the system and the long timescale could facilitate a 
first order transition. Such a transition could be modeled through a bubble 
nucleation scenario and Witten\cite{rf:10} argued that the trapped false 
vacuum domains (TFVD) ({\it i.e.} the quark phase) could contain a substantial 
amount of baryon number. QCD-motivated studies\cite{rf:11,rf:12} of baryon 
evaporation from such TFVD (called strange quark nugget or SQN hereafter)
showed that if the SQNs contain baryon number in excess of 10\(^{40-42}\),
they would be stable on cosmological time scales and be viable candidates
for CDM, as they would be extremely non-relativistic. Note that the  
number of baryons (which would subsequently take part in BBN) within the
event horizon at the microsecond epoch is 10\(^{49-50}\); thus the baryon 
number contained in SQNs could be 3-4 times as much and SQNs of size 
\(\ge \) 10\(^{42}\) could be easily accommodated. However, it has to be 
reiterated over and over again that the baryon number contained in the SQNs 
is in the form of quarks and they do not participate in BBN at all. Thus, to 
distinguish them from the usual baryons and yet account for the baryon number 
contained in them, we coin the term {\it quasibaryonic} to describe the SQNs 
and the CDM component coming from SQNs as {\it quasibaryonic dark matter}, 
distinct from nonbaryonic or baryonic dark matter.
\par
We can estimate the size of the SQNs formed in the {\it first order} 
cosmic QCD transition in the manner prescribed by Kodama, Sasaki 
and Sato\cite{rf:13} in the context of the GUT phase transition. 
For the sake of brevity, let us recapitulate very briefly the salient 
points here; for details, please see Alam {\it et al}\cite{rf:14} and 
Bhattacharyya {\it et al}\cite{rf:15}. Describing the cosmological
scale factor \( R \) and the co-ordinate radius \( X \) in the 
Robertson-Walker metric through the relation
\begin{equation} 
ds^2 = -dt^2 + R^2 dx^2 = -dt^2 + R^2\{dX^2 + X^2(sin^2 \theta d\phi^2 
+ d\theta^2)\}, 
\end{equation}
one can solve for the evolution of the scale factor \( R(t) \) in the 
mixed phase of the first order transition. In a bubble nucleation description 
of the QCD transition, hadronic matter starts to appear as individual bubbles 
in the quark-gluon phase. With progressing time, they expand, more and more 
bubbles appear, coalesce and finally, when a critical fraction of the total 
volume is occupied by the hadronic phase, a continuous network of hadronic 
bubbles form (percolation) in which the quark bubbles get trapped, the TFVDs. 
The time at which this happens is the percolation time \( t_p \), whereas 
the time when the phase transition starts is denoted by \( t_i \). Then, the 
probability that a spherical\footnote{In general, the TFVDs need not be 
spherical. For the QCD bubbles, however, it is believed that there is a 
sizable surface tension which would facilitate spherical bubbles.} region 
of co-ordinate radius \( X \) lies entirely within the quark bubbles would 
obviously depend on the nucleation rate of the bubbles as well as the 
coordinate radius \( X(t_p,t_i) \) of bubbles which nucleated at \( t_i \) 
and grew till \( t_p \). For a nucleation rate \( I(t) \), this probability 
\( P(X,t_p) \) is given by 
\begin{equation}
P(X,t_p) = exp \left[-\frac{4 \pi}{3}\int_{t_i}^{t_p} dt I(t) R^3(t) [X +
X(t_p,t_i)]^3\right].
\end{equation}
After some algebra,\cite{rf:15} \ it can be shown that if all the CDM is 
believed to arise from SQNs, then their size distribution peaks, for 
reasonable nucleation rates, at baryon number \( \sim \) 10\(^{42 - 44}\), 
evidently in the stable sector. Recalling that \(\Omega_{B} \sim \) 0.1 
corresponds to 10\(^{49-50}\) baryons within the horizon at the microsecond
epoch, the total baryon number contained in SQNs to account for 
\(\Omega_{CDM} \sim \) 0.35 would imply 10\(^{7-9}\) SQNs within the horizon 
just after the QCD phase transition. Because of their enormous mass, they 
would be nonrelativistic immediately after their formation. It may
thus be remarked that they would be discrete macrosopic bodies (radius 
\( R_N \sim \) 1m) separated by rather large distances (100 - 300m) in the 
background of the radiation fluid. 
\par
Any deviation from a uniform distribution of SQNs should result in a large 
attractive force, under which they should gravitate toward one another. Given 
the property that they become more and more bound with increasing 
mass,\cite{rf:10} \ they should tend to coalesce and grow to larger sizes. 
However, the radiation pressure acting on the moving SQNs would serve to 
inhibit such motion till such time when the gravitational force dominates 
over it. We can estimate the relative magnitude of these two forces in a 
straightforward manner. If the number of SQNs within the horizon at the time 
\( t_p \) is 10\(^9\) (see above), then the number at any later temperature 
\( T \) is given by
\begin{equation}
N_N(T) = 10^9 \left(\frac{100 \ MeV}{T}\right)^3
\end{equation}
and the number density of SQNs is consequently
\begin{equation}
n_N(T) \equiv \frac{N_N}{V_H} = \frac{3 N_N}{4 \pi (2t)^3}
\end{equation}
since the horizon length in the radiation dominated era is \( 2 t \). The 
time \( t \) and temperature \( T \) are related by the relation
\begin{equation}
t = 0.3 g_*^{-1/2} \frac{m_{pl}}{T^2}
\end{equation}
with \( g_* \sim \) 17.25 after the QCD transition.\cite{rf:14}
\par
The expression for the gravitational force as a function of temperature
\( T \) can be written as 
\begin{equation}
F_{\mathrm{grav}}=\frac {G M_N^2 } {{\bar{r}_{nn}(T)}^2}
\end{equation}
where \( M_N \) is the SQN mass. (For the sake of simplicity, we assume
that all SQNs have the same mass.)  \( \bar{r}_{nn}(T) \) is the mean 
separation between the two nuggets, estimated from the density of nuggets 
at the temperature \( T \). The force due to the radiation pressure on the 
SQNs arises only due to their relative motion; when they are at rest, there
is no resultant force on them. But when the SQNs are in motion, the radiation
fluid in front of the moving SQN gets compressed and thus exerts an 
additional pressure opposing the motion. In a relativistic 
framework,\footnote{Even though the SQNs are nonrelativistic, a relativistic 
treatment is necessary to handle the radiation fluid.} \ this amounts to  
a force \( F_{\mathrm{rad}} \) given by
\begin{equation}
F_{\mathrm{rad}}=\frac{1}{3} \rho_{\mathrm{rad}} c v_{\mathrm{fall}}
(\pi R_N^2) \beta \gamma
\end{equation}
where \( \rho_{\mathrm{rad}} \) is the total radiation energy density, 
counting all relativistic species at the temperature \( T \), 
\( v_{\mathrm{fall}} \) (or \( \beta c \)) is the velocity of the SQN and
\( \gamma \) the corresponding Lorentz factor. The ratio of these two forces,
\( F_{\mathrm{grav}}/F_{\mathrm{rad}} \) is plotted against temperature 
in Fig.~1 for SQNs of initial size 10\(^{42} \). 
\par
\begin{figure}
\epsfxsize = 15cm
\centerline{\epsfbox{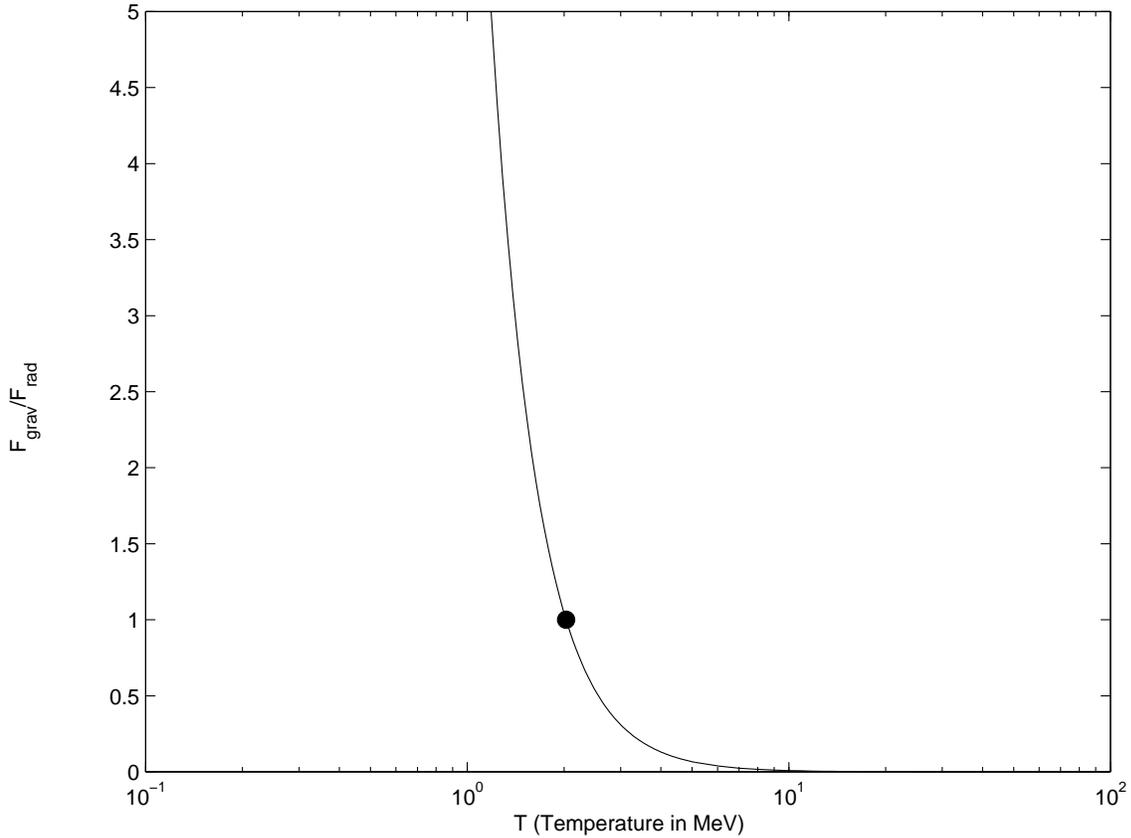}}
\caption{Variation of \( F_{\mathrm{grav}} / F_{\mathrm{rad}} \)
with temperature. The dot represents the point where the ratio assumes the
value 1.}
\label{fig1}
\end{figure}
\par
Fig.~1 readily reveals that the ratio \( F_{\mathrm{grav}}/F_{\mathrm{rad}} \)
is very small initially. As a result, the nuggets will remain separated due to
the radiation pressure. For temperatures lower than a critical value 
\( T_{\mathrm{cl}} \), the gravitational force starts dominating,
facilitating the coalescence of the SQNs under mutual gravity. It should 
come as no surprise that even for very small values of \( \beta \), the large
surface area of the SQN is responsible for a considerable resisting force due
to radiation pressure.
\par
We can estimate the size of the coalesced SQNs from the number of SQNs within
the horizon at \( T_{cl} \). This is, of course, a lower limit, as the 
collapse would only start at \( T_{cl} \) and would take some time during 
which more SQNs will enter the horizon. For SQNs of size 10\(^{42}\), the 
total mass in quasibaryonic matter within the horizon at \( T_{cl} \) 
turns out to be \( \sim \) 0.12\( M_{\odot} \); for 10\(^{44}\), it is 
0.01\( M_{\odot}\). The actual value could be much (3-10 times) higher but 
that can be ascertained only through a detailed simulation. Such a calculation 
is rather involved and remains a future project. In any case, it
can be safely assumed that the coalesced SQNs will have masses above the 
fusion threshold of 0.08\(M_{\odot}\) and  once coalesced, the density of 
the resulting configuration would be so low that they cannot clump any 
further. (It may also be recalled at this point that quasibaryonic blobs
cannot grow to arbitrarily large sizes. We have earlier shown\cite{rf:16} \ 
that there exists an upper limit on the mass of astrophysical compact 
quark matter objects, about 1.4\(M_{\odot}\) for the canonical value of 
(145 MeV)\(^4\) for the MIT Bag parameter.) They could thus persist till 
the present time and manifest themselves as MACHOs. For \(\Omega_{CDM} \sim\) 
0.35, there would be about 10\(^{23-24}\) such objects within the horizon 
today and about 2 - 3 \(\times\) 10\(^{13}\) within the Milky Way halo. We 
should verify whether this abundance is consistent with the observed number 
of MACHO events. 
\par
The abundance of gravitational lenses is estimated through the optical
depth which is the probability that a given star lies within the Einstein
ring of a lens, {\it i.e.} the number density of the lenses times the area
of the Einstein ring of each lens. The expression reads\cite{rf:17}
\begin{equation}
\tau = \frac{4 \pi G}{c^2} D_{s}^{2}\int \rho(x) x (1-x) dx 
\end{equation}
where \( D_s \) is the distance between the observer and the source
(a star in the LMC in the present case) and \( x=D_{d} {D_{s}}^{-1} \), 
\( D_d \) being the distance between the observer and the lens. In particular, 
\( \rho \) is the mass-density of the MACHOs, which is of the form 
\( \rho = \rho_0 \frac{1}{r^2} \) in the spherical halo model. Assuming a
halo extending all the way upto the LMC, we obtain \( \tau \simeq \) 
2 - 5\(\times\)10\(^{-7}\), in excellent agreement with observation.\cite{rf:5}
\par
The scenario proposed here could have important consequences. The presence 
of MACHOs prior to BBN could act as a large sink for baryons (primarily 
neutron) and influence inhomogeneous BBN significantly. This is now under 
investigation.\cite{rf:18} \ Also the gravitational attraction exerted by the
MACHOs on the surrounding matter could act as a seed of inhomogeneity for
galaxy formation processes. This will have to be explored in depth. 
Furthermore, the existence of an upper mass limit for astrophysical quark 
objects implies that if two or more MACHOs collide and merge, the resultant 
body may exceed this upper limit. In such cases, the additional mass would 
have to be shed and these could be the source of the ultrahigh energy 
(\(\ge\) 10\(^{20}\) eV) cosmic rays, without the necessity of invoking an
acceleration mechanism.
\par
To conclude, we have shown that in a first order cosmic quark-hadron 
phase transition, quasibaryonic dark matter could arise, entirely within
the framework of the standard model of particle interactions, which would 
explain all of CDM. The observed halo MACHOs could be the natural 
manifestation of quasibaryonic CDM.



\begin{thebibliography}{99}
\bibitem{rf:1}
C.~Alcock {\it et al}, \JL{Nature,365,1993,621}
\bibitem{rf:2}
E.~Aubourg {\it et al}, \JL{Nature,365,1993,623} 
\bibitem{rf:3}
B.~Paczynski, \JL{Ap.\ J.,304,1986,1}
\bibitem{rf:4}
W.~Sutherland, \JL{Rev.\ Mod.\ Phys.,71,1999,421}
\bibitem{rf:5}
C.~Alcock {\it et al}, \JL{Ap.\ J.,542,2000,281}
\bibitem{rf:6}
D.~N.~Schramm, in : {\it Proc. ICPA-QGP'97{\rm :} Physics and Astrophysics 
of Quark-Gluon Plasma}, (Eds. B.~Sinha, Y.~P.~Viyogi and D.~K.~Srivastava),
p.29, (Narosa Publishing, New Delhi, 1998)
\bibitem{rf:7}
K.~Jedamzik and J.~C.~Niedemeyer, \PR{D59,1999,124014}
\bibitem{rf:8}
C.~Schmid, D.~J.~Schwarz and P.~Widerin, \PR{D59,1999,043517} 
\bibitem{rf:9}
J.~Alam, S.~Raha and B.~Sinha, \JL{Phys.\ Rep.,273,1996,243}
\bibitem{rf:10}
E.~Witten, \PR{D30,1984,272}
\bibitem{rf:11}
P.~Bhattacharjee, J.~Alam, B.~Sinha and S.~Raha, \PR{D48,1993,4630}
\bibitem{rf:12}
K.~Sumiyoshi and T,~Kajino, \JL{Nucl.\ Phys.\ (Proc.\ Suppl.),24,1991,80}
\bibitem{rf:13}
H.~Kodama, M.~Sasaki and K.~Sato, \PTP{68,1982,1979}
\bibitem{rf:14}
J.~Alam, S.~Raha and B.~Sinha, \JL{Ap.\ J.,513,1999,572}
\bibitem{rf:15}
A.~Bhattacharyya {\it et al}, \PR{D61,2000,083509}
\bibitem{rf:16}
S.~Banerjee, S.~K.~Ghosh and S.~Raha, \JP{G26,2000,L1}
\bibitem{rf:17}
R.~Narayan and M.~Bartelmann, in : {\it Formation of structure in the 
universe} (Eds. A.~Dekel and J.~P.~Ostriker), (Cambridge University 
Press, Cambridge, UK, 1999)
\bibitem{rf:18}
T.~Kajino {\it et al}, under investigation.
\end{thebibliography}
\end{document}